\documentclass[usenatbib]{mnras}
\usepackage{epsfig}
\usepackage{amsmath,amssymb}
\usepackage{color}
\usepackage{booktabs}
\usepackage{pdflscape}
\usepackage{textcomp}

\usepackage{epsfig}
\usepackage{color}
\usepackage{soul}
\usepackage{caption}
\usepackage{aas_macros}  
\usepackage{chngcntr}
\graphicspath{{figures/}}

\newcommand{\Msolar}{\mbox{\,$\rm M_{\odot}$}}        

 \newcommand{\teff}{\mbox{\,$T_{\rm eff}$}}      

  \def\simge{\mathrel{\raise1.16pt\hbox{$>$}\kern-7.0pt
    \lower3.06pt\hbox{{$\scriptstyle \sim$}}}}           
  \def\simle{\mathrel{\raise1.16pt\hbox{$<$}\kern-7.0pt
    \lower3.06pt\hbox{{$\scriptstyle \sim$}}}}           

\title[K2 observations of HD144941]{K2 spots rotation in the helium star HD144941}

\author[C. S.~Jeffery \& G.~Ramsay]{C. Simon Jeffery$^{1,2}$\thanks{email: csj@arm.ac.uk} and Gavin Ramsay$^1$ \\
$^{1}$Armagh Observatory and Planetarium, College Hill, Armagh BT61 9DG, UK\\
$^{2}$School of Physics, Trinity College Dublin, College Green, Dublin 2, Ireland
}
\begin{document}

\date{Accepted 2018 January 30. Received 2018 January 30; in original form 2018 January 10}


\maketitle

\label{firstpage}

\begin{abstract}

  HD144941 is an evolved early-type metal-poor low-mass star with a
  hydrogen-poor surface.  It is frequently associated with other
  intermediate helium-rich subdwarfs and extreme helium stars.
  Previous photometric studies have failed to detect any variability.
  New observations with the {\it K2} mission show complex but periodic
  variations with a full amplitude of 4 parts per thousand.  It is
  proposed that these are due to an inhomogeneous surface brightness
  distribution (spots) superimposed on a rotation period of
  $13.9\pm0.2$\,d.  The cause of the surface inhomogeneity is not
  identified, although an oblique dipolar magnetic field origin is
  plausible.
\end{abstract}


\section{Introduction}
\label{s:intro}

HD144941 was reported helium-rich by \citet{macconnell70}.  The
spectrum was shown to be metal-poor with a helium/hydrogen ratio
$\approx 10$
\citep{hunger73,harrison97,jeffery97,beauchamp97,przybilla05,pandey17}.
Its effective temperature and surface gravity are appropriate for a
main-sequence B star, but it is 6 magnitudes too faint to be
associated with the Sco-Cen OB2 association within which it lies
\citep{macconnell70}.  It has subsequently been compared with extreme
helium stars, including the pulsating helium star V652\,Her
\citep{jeffery01a}, which is known to be a low-mass evolved star.

Given their power for yielding distance-independent radii and masses,
\citet{jeffery96c} sought evidence for pulsations in HD144941.  Their
failure to find light variations with semi-amplitudes $>0.0053$
mag. on timescales between 8 hours and 50 days, or with
semi-amplitudes $>0.0023$ mag. on timescales between 6 minutes and 5
hours, was important.  Like V652\,Her, HD144941 has a temperature and
radius within the instability zone for pulsations driven by an opacity
bump associated with iron-group elements.  Pulsation instability is
enhanced by a deficiency in hydrogen, but reduced by a deficiency in
metals \citep{saio93,jeffery99,jeffery16a}.  Thus the low metallicity
in HD144941 is consistent with the absence of pulsations.  Taken
together, V652\,Her and HD144941 confirmed the r\^ole of metals in
driving pulsations in early-type helium stars.

A subsequent space-based photometric study of HD144941 using the {\it
  STEREO} satellites failed to detect any variability
\citep{wraight12}.

HD144941 = EPIC 203109319 was observed during the {\it Kepler}
follow-up mission {\it K2} and reported to show ``other periodic or
quasiperiodic variability" with a period of $\approx 7$ d
\citep{armstrong16}.  Given the unusual properties of HD144941, the
question arises as to the nature and cause of this variability.  This
paper presents the light curve (\S\,2), examines its form in more
detail (\S\,3), and discusses the origin and implications of its
variability (\S\,4).

\begin{figure}
\epsfig{file=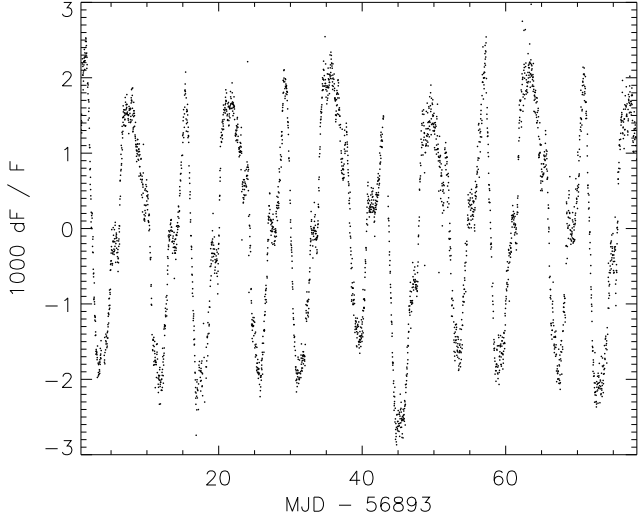,height=95mm,clip=,angle=90}
\caption{The {\it K2} light-curve of HD144941 made between 2014 Aug 23
  and 2014 Nov 10. The cadence is one photometric point per 29.4
  min. We have normalised it so that the mean count rate is zero and
  the deviations are given in parts per thousand. We have also lightly
  detrended the light curve to remove residual long term variations.}
\label{f:light}
\end{figure}

\section{Observations}
\label{s:obs}

HD 144941 was selected as a target for {\it K2} \citep{howell14} in
Campaign 2 by Papics (GO 2010) as part of a wider project to study
single B-type stars.  It was observed in long cadence mode (one
photometric point per 29.4 min) without significant interruption
between 2014 Aug 23 and 2014 Nov 10 (MJD=56893.8--56971.3).  During
the course of a {\it K2} campaign, tiny adjustments are made to the
spacecraft pointing.  A number of groups have developed tools to
mitigate the effects of these adjustments and have made `cleaned'
photometry publicly available.  We have taken the light curve of HD
144941 as extracted by \citet{vanderburg14}.  To remove any residual
long term trends in the light curve we detrended using a 3rd order
polynomical.  This light curve is shown in Fig.~\ref{f:light}.

\begin{figure}
\epsfig{file=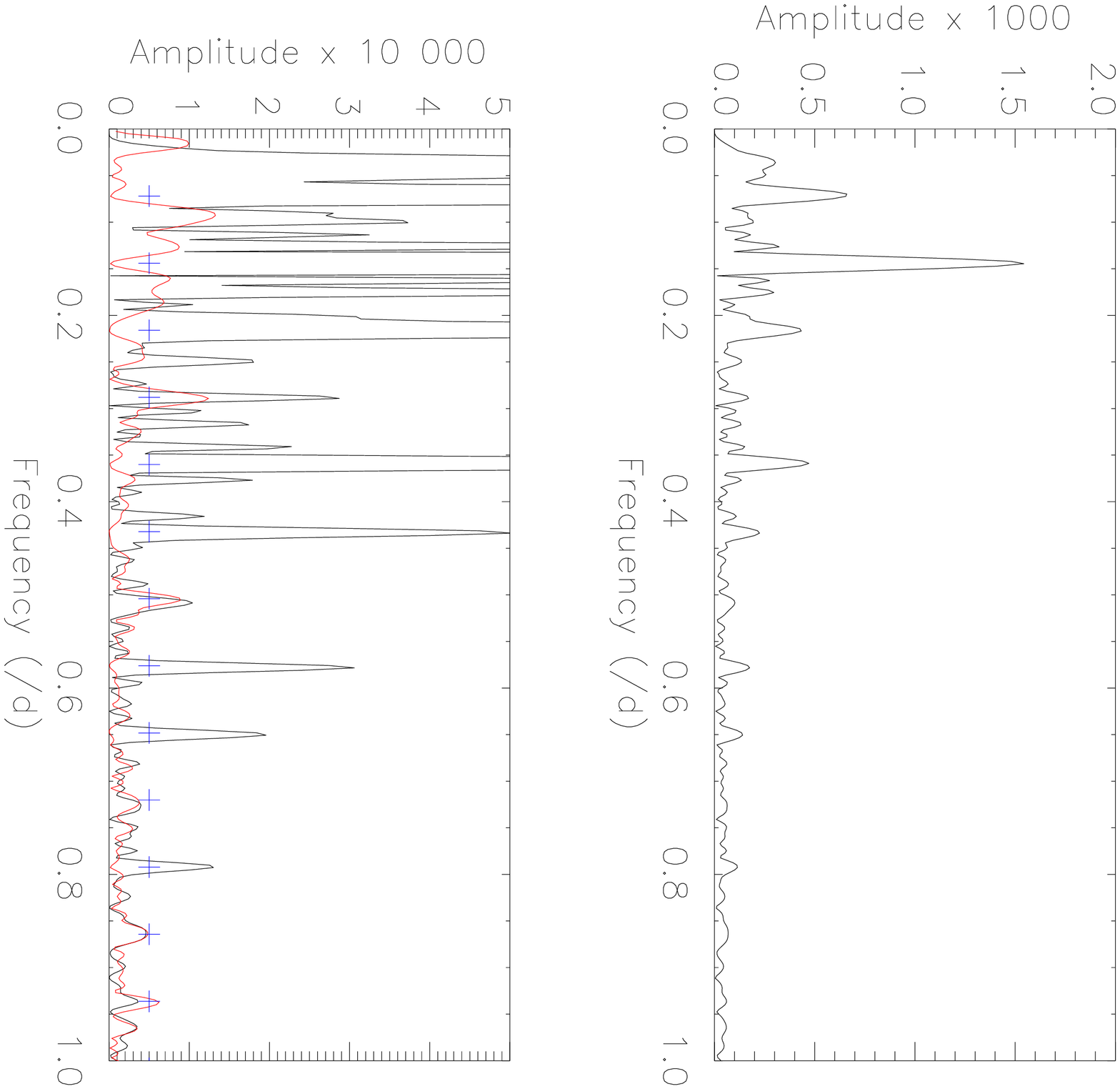,height=95mm,clip=,angle=90}
\caption{The amplitude spectrum of the {\it K2} light-curve of
  HD144941. Top and bottom: the amplitude spectrum of the complete
  data (black). Bottom: data prewhitened (red) by nine strongest
  harmonics and a contribution from low-frequency noise, together with
  locations of the fundamental period and eleven harmonics (blue
  crosses). }
\label{f:power}
\end{figure}

\begin{figure}
\epsfig{file=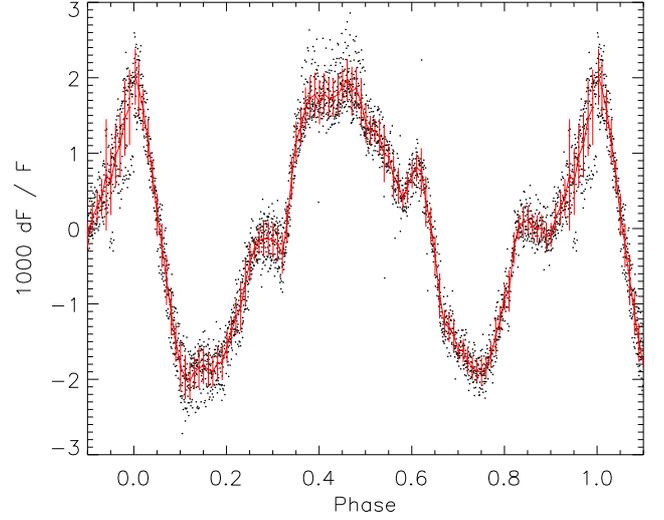,height=95mm,clip=,angle=90}
\caption{The {\it K2} light-curve of HD144941 folded on a period of
  13.93 d (black dots) and binned (red symbols; error bars are
  standard deviations of data within each phase bin). }
\label{f:phase}
\end{figure}

\begin{table}
\begin{center}
\caption{Fourier decomposition of {\it K2} light curve of HD144941.}
\label{t:fourier}
\begin{tabular}{rrr}
\hline
$n$ & $f$ (d$^{-1}$) & $a$ (\textperthousand)  \\
\hline
1 & 0.07180(0613)& 0.636 \\ 
2 & 0.14403(0216) & 1.531 \\ 
3 & 0.21600(0778)  & 0.454  \\ 
4 & 0.28814(2298) & 0.142  \\
5 & 0.35994(0700) & 0.459  \\
6 & 0.43217(1961) & 0.163  \\ 
8 & 0.57655(1918)  & 0.167  \\
9 & 0.64886(2471) & 0.129   \\ 
11 & 0.79237(3024)  & 0.105 \\[2mm]
-- & 0.03240(1227) & 0.284  \\ 
-- & 0.05167(1754) & 0.218  \\
\hline
\end{tabular}
\end{center}
\end{table}

\section{Analysis}

The {\it K2} light-curve in Fig.~\ref{f:light} suggests an apparent
period of $\approx 7$ d. Closer inspection shows that alternate cycles
are more similar than consecutive cycles.  Fourier analysis
(Fig.~\ref{f:power}) demonstrates a highly-structured singly-periodic
signal, with a fundamental period of $13.9\pm0.2$\,d and multiple
harmonics.  Some low frequency power is also present, which is
probably red noise.  Prewhitening the data by the fundamental period,
8 harmonics and a red-noise contribution gives the residual amplitude
spectrum shown in Fig.~\ref{f:power}. The adopted frequencies ($f$ in
cycles per day) and amplitudes ($a$ in parts per thousand -
\textperthousand) are shown in Table~\ref{t:fourier}, with harmonics
labelled as $n$.  Detrending the red-noise signals allows the data to
be phase folded, whilst phase binning at intervals of 0.01 cycles
brings the light curve structure into sharper focus
(Fig.~\ref{f:phase}).

Defining zero phase by the sharper of the two light maxima, the
13.9\,d cycle shows two similar minima at phases 0.2 and 0.75, and
shoulders on the ascending branches at phases 0.3 and 0.85. The two
maxima are quite different in character. The segments which appear
most similar are separated by $\approx0.6$ cycles and not an exact
half cycle.

\section{Interpretation}

This is the first high-definition light curve obtained for {\it any}
strongly hydrogen-deficient star, albeit one having 10\% hydrogen on
its surface.  At 4\textperthousand, the full-amplitude of the
variation is less than half the lower limit for periods longer than
one third of a day measured by \citet{jeffery96c}.  The period of
13.9\,d is a factor of 100 too long to associate with radial
pulsations such as seen in the comparable extreme helium star
V652\,Her, or in main-sequence $\beta$ Cepheid variables
\citep{aerts10}. A non-radial g-mode pulsation is unlikely on similar
grounds. In either case, the highly-structured light curve is not
anticipated from a small-amplitude pulsation, where a more sinusoidal
form is anticipated.

It is proposed that the light curve in HD144941 is due to small-scale
structure (or spots) on the stellar surface which rotates with a
period of 13.9\,d.  This structure is dominated by a single dipole
distribution, superimposed by small scale features which produce sharp
jumps in total brightness as they appear onto or disappear from the
visible hemisphere.

The difficulty is that HD144941 is a B-type star with an effective
temperature of some 23\,000\,K \citep{harrison97}.  Since the surface
layers should be completely radiative, it is difficult to explain how
such structures should arise.  A possible corollary is provided by the
surfaces of other chemically peculiar B stars, notably the Bp(He)
stars such as $\sigma$ Ori E \citep{greenstein58,hesser76,walborn76},
where a strong magnetic field ($\approx7.5$ kG, \citep{oksala15})
modifies the local surface structure and emergent fluxes.  Such stars
show strong variations in light (0.2 mag in $u$, 0.1 mag in the {\it
  MOST} filter for $\sigma$ Ori E) over the course of the rotation
period (1.19\,d) \citep{townsend13}.  It is interesting that the
HD144941 light curve is more structured than that of $\sigma$ Ori E as
measured from space with the {\it MOST} spacecraft \citep{townsend13}.
With a much smaller amplitude ($\times0.04$), HD144941 need only
possess a weak, if complex, magnetic field in order to exhibit
inhomogeneities sufficient to explain the {\it K2} light curve.  If
light amplitude scales with field strength, then a polar field of some
300\,G would be sufficient, and detectable.

The question then becomes whether spots induced by magnetic fields are
present on other evolved chemically peculiar stars.  A magnetic-field
origin for multi-periodic variability in the hydrogen-weak subdwarf LS
IV-14 116 was suggested by \citet{naslim11}, and ruled out by
\citet{green11} and \citet{randall15}.  Long-term variability observed
in the hydrogen-weak subdwarfs KIC1044976 \citep{jeffery13a} and UVO
0825+15 \citep{jeffery17a} has still to be explained.

An alternative line of thought has suggested that HD144941 might be a
main-sequence Bp(He) star and not an evolved star, although the low
metallicity ([Fe/H] $= -1.9\pm0.2$ \citet{jeffery97}) argues strongly
against such a view.  Given the apparent magnitude ($m_{\rm
  V}=10.14\pm0.01$), extinction ($E_{\rm B-V}=0.25\pm0.02$
\citep{jeffery86.iue}), effective temperature ($\teff =
22\,000\pm1000$\,K) and surface gravity ($\log g = 4.15\pm0.1$
\citep{przybilla05}), only an accurate distance is necessary to
estimate the mass.  The TGAS survey provides a parallax for HD144941 =
TYC 6788-284-1 of $0.00101\pm0.00061$" \citep{astraatmadja16,gaia16b}
which translates to a mass of $1.2\pm0.7$\Msolar. Although slightly
large for the evolved star argument, this rules out the main-sequence
star argument unless the distance is substantially increased in the
second {\it Gaia} data release.

\section{Conclusion}

HD144941 has hitherto been known as a non-variable low-mass helium
star with comparable dimensions to the pulsating helium star
V652\,Her.  Its low metallicity accounts for the absence of
pulsations, despite lying in the Z-bump instability strip.

Photometric observations with {\it K2} radically challenge this
picture, demonstrating a 13.9\,d light curve which can best be
explained by the rotation of a star with an inhomogeneous surface
brightness.  Explaining the origin of the surface inhomogeneity
presents major challenges. It will at least require spectroscopic
evidence for a structured surface and a magnetic field on HD144941,
data on the long-term behaviour of the light curve and evidence for
similar phenomena on related hydrogen-deficient stars.

Whilst an alternative view of HD144941 as a chemically peculiar
main-sequence star is convenient, the TGAS parallax measurement
supports the argument for a low-mass evolved star.  The {\it K2}
observation has implications for interpreting the properties of other
evolved hydrogen-deficient stars, including both extreme helium stars
and hot subdwarfs; some of these must become targets for future space
photometry missions.

\section*{Acknowledgments}

This paper includes data collected by the {\it Kepler}
mission. Funding for the {\it Kepler} mission is provided by the NASA
Science Mission directorate. We thank Andrew Vanderburg for making his
light curves publically available.

This work has made use of data from the European Space Agency (ESA)
mission {\it Gaia} (\url{https://www.cosmos.esa.int/gaia}), processed
by the {\it Gaia} Data Processing and Analysis Consortium (DPAC,
\url{https://www.cosmos.esa.int/web/gaia/dpac/consortium}).  Funding
for the DPAC has been provided by national institutions, in particular
the institutions participating in the {\it Gaia} Multilateral
Agreement.

The Armagh Observatory and Planetarium is funded by direct grant form
the Northern Ireland Dept for Communities.  The authors acknowledge
support from the UK Science and Technology Facilities Council (STFC)
Grant No. ST/M000834/1.

\bibliographystyle{mn2e}
\bibliography{hd144941}

\label{lastpage}
\end{document}